# Structural, dielectric, ferroelectric and tuning properties of Pb-free ferroelectric $Ba_{0.9}Sr_{0.1}Ti_{1-x}Sn_xO_3$

H. Zaitouni[1,*], L. Hajji[1], D. Mezzane[1], E. Choukri[1], Y. Gagou[2], K. Hoummada[3], A. Charai[3], A. Alimoussa[1], B. Rožič[4], M. El Marssi[2], Z. Kutnjak[4]

[1]IMED-Lab, Cadi-Ayyad University, Faculty of Sciences and Technology, Department of Applied Physics, Marrakech, Morocco.

[2]Laboratory of Physics of Condensed Matter (LPMC), University of Picardie Jules Verne, Scientific Pole, 33 rue Saint-Leu, 80039 Amiens Cedex 1, France.

[3]Laboratoire Matériaux et Microélectronique Nanosciences de Provence, Aix Marseille Université-CNRS, Faculté des Sciences, Marseille, France

[4]Laboratory for calorimetry and dielectric spectroscopy, Condensed Matter Physics Department, Jozef Stefan Institute, Ljubljana, Slovenia.

**Abstract**

A series of Pb-free ferroelectric materials $Ba_{0.9}Sr_{0.1}Ti_{1-x}Sn_xO_3$ (BSTS-x) with $0 \leq x \leq 0.15$ was successfully prepared *via* solid-state reaction method. The effect of Sn substitution on the crystal structure, microstructure, dielectric behavior, ferroelectric and tunable features of BSTS-x ceramics were systematically investigated. Room temperature (RT) x-ray diffraction (XRD) analysis using the Rietveld refinement method reveals that all the synthesized BSTS-x ceramics were well crystallized into single perovskite structure. The results show a tetragonal phase for $0.00 \leq x \leq 0.02$, which evolves to orthorhombic and tetragonal coexisting phases for $0.05 \leq x \leq 0.07$. The composition $x = 0.10$ showed a mixture of tetragonal, orthorhombic and rhombohedral phases at RT, while a single cubic phase is observed for $x = 0.15$. The crystal phases determined by XRD were confirmed by Raman spectroscopy. Enhanced dielectric permittivity with a maximum value of $\varepsilon' \sim 35000$ is observed for $x = 0.10$ at RT. The ferroelectric behavior of BSTS-x ceramics was investigated through polarization hysteresis loops and tunability measurements. High tunability of 63% at RT and under the low DC-applied electric field of 1.40 kV/cm is achieved for $x = 0.10$.

* Corresponding author: hajar.zaitouni@ced.uca.ma; hajarzaitouni88@gmail.com

## 1. Introduction

During the last few decades, lead free-ferroelectric materials caught the attention of researchers resulting in numerous studies. Due to its non-toxicity, barium titanate (abbreviated BT) ceramic is an attractive candidate to replace lead-based materials. Its dielectric, phase transition temperature and ferroelectric properties can be tailored by suitable doping the perovskite structure, either in the square site (occupied by Ba) or the octahedral one (occupied by Ti). Among this family, barium strontium titanate (BST) with high dielectric constant combined with low dielectric loss attracts a particular interest for dynamic random-access memory (DRAM) applications and multilayer ceramic capacitors [1–5]. The BST system is well recognized by its strong response to the DC bias electric field. This characteristic is desirable for tunable microwave devices like phase shifters and tunable filters, etc. [5–7]. Moreover, the isovalent substitution of $Ba^{2+}$ by $Sr^{2+}$ is also known to shift the Curie temperature from high to lower temperatures only by controlling the molar fraction of $Ba^{2+}/Sr^{2+}$.

It is worth to note that the substitution of $Ti^{4+}$ by $Sn^{4+}$ in BT lattice (abbreviated BTS) has also been an effective way to shift the Curie temperature close to room temperature and also to induce various interesting properties in the dielectric behavior and sensor applications [8–10]. Besides, BTS solid solutions have attracted significant attention for their potential applications in electric-field tunable devices [11]. Detailed examinations of dielectric and ferroelectric properties depending on Sn amount in BTS have been investigated in the literature. However, the literature survey shows limited reports on the effect of substitution of $Sn^{4+}$ in BST host lattice [12, 13]. Thus, the present paper aims to examine the $Ba_{0.9}Sr_{0.1}Ti_{1-x}Sn_xO_3$ ceramic family by performing a complete study on its crystal structure, dielectric and ferroelectric properties, including the DC-electric field dependence of the dielectric constant. A series of $Ba_{0.9}Sr_{0.1}Ti_{1-x}Sn_xO_3$ ceramics, with x = 0.00, 0.02, 0.05, 0.07, 0.10 and 0.15 (denoted BSTS-x) was fabricated *via* the conventional solid-state reaction method. A detailed structural refinement analysis based on the Rietveld method was performed on RT x-ray (XRD) patterns. Raman spectroscopy and Scanning Electron Microscope (SEM) measurement permitted to confirm phase transitions and the symmetries. Using dielectric spectroscopy, P-E and capacitance-voltage (C-V) hysteresis measurements, we determined the value of saturated polarization and highlighted strong ferroelectric activity in this system.

## 2. Experimental procedure

The polycrystalline BSTS-x ceramics with (x=0, 0.02, 0.05, 0.07, 0.10 and 0.15) were synthesized by the conventional solid-state reaction method. The basic materials are $BaCO_3$, $SrCO_3$, $TiO_2$ and $SnO_2$ powders that were weighed stoichiometrically and mixed with ethanol in an agate mortar for 2h. The mixed powder was dried and then calcined at 1150°C for 12h in air. After the addition of 5 wt.% polyvinyl alcohol (PVA) to the calcined powder as a binder, the powders were pressed into pellets of approximately 13 mm diameter and 1-2 mm thickness using the uniaxial hydraulic press. The pressed pellets were heated up to 700°C to remove the binder and then sintered at 1350-1400 °C, depending on Sn amount.

The XRD patterns were recorded using PANalytical EMPYRAN diffractometer under CuKα radiation ($\lambda_{K\alpha1}$ = 1.5405 Å and $\lambda_{K\alpha2}$ = 1.5444 Å) at room temperature. Rietveld analysis of XRD data was performed by using the FullProf suite software. By using the TESCAN VEGA3 Scanning Electron Microscope (SEM), the grain morphology of the sintered pellets was observed and its relation to the Sn-amount established. At the same time, the ceramics bulk density was measured by the Archimedes' method. Raman spectroscopy measurements of the sintered ceramics were performed to support the XRD study by using a micro-Raman RENISHAW with a CCD detector and green laser excitation of 532 nm. Dielectric measurements were conducted in the frequency range of 1 Hz – 1 MHz and temperature interval from -120 °C to 250 °C by the Solartron Gain phase Impedance Analyzer SI-1260. Temperature-dependent P-E hysteresis loops at 5 Hz and room temperature C-V profiles were measured by using a ferroelectric test system (aixACCT, TF Analyser 3000).

## 3. Results and discussion
### 3.1 X-ray diffraction analysis

RT x-ray diffraction (XRD) patterns of BSTS-x ceramics are shown in Fig. 1(a). All the patterns show reflections peaks typical to $BaTiO_3$ perovskite structure with no traces of impurities at the limit of device sensitivity, suggesting that $Sn^{4+}$ has successfully diffused into the host lattice. The enlarged (200) reflection peak in the higher Sn-amount composition, around $2\theta = 45°$, as shown in Fig. 1(b) is consistent with the coexistence of structural phases in intermediate compositions from tetragonal to cubic phases. Moreover in Fig.1(b), it can be seen that in BSTS-x ceramics with x = 0.00 and 0.02 (200) reflection peak splits into (200) and (002) peaks that are fingerprints of tetragonal (T) phase, which is similar to previous reports [13, 14]. For $0.05 \leq x \leq 0.07$, the phase structure shifts from the (T) phase to the coexistence of Orthorhombic (O) and (T) phases. When x increases to 0.10, mixed diffraction peaks are observed, suggesting the coexistence of Rhombohedral (R), (T) and (O) phases. From x = 0.15 pure Cubic (C) phase is observed at RT. These results show a sensitive dependence of the structural symmetry as a function of Sn-amount. Moreover, Fig.1(b) shows the splitting of the diffraction lines at higher (2θ) angles due to the presence of K$\alpha_1$ and K$\alpha_2$ radiation in x-ray source. Furthermore, it can be seen that all the diffraction peaks shift to low angles (2θ) (Fig. 1(b)) with increasing Sn-content, which is due to the partial replacement of the Ti$^{4+}$ ions with small radius (r = 0.605 Å) by the Sn$^{4+}$ ions with larger radius (r = 0.690 Å).

**(Insert Fig. 1, here)**

Rietveld refinement of recorded x-ray patterns in BSTS-x ceramics with (x = 0.00 to 0.15) was also conducted. The calculations in profile adjustments mode lead to a rapid convergence for all the compositions. From these results, the structural model was established by using the Rietveld technique based on the least square calculation taking into account atomic positions, occupation rate and thermal agitation. The crystal structure of the compositions with x = 0.00 and 0.02 were successfully refined in the (T) symmetry of the *P4mm* space group, while the composition with x = 0.15 was refined in the (C) symmetry of the *Pm-3m* space group. Optimal reliability factor was then observed satisfactorily. The refinement of BSTS-x samples (x = 0.05 and 0.07) using only a single T (*P4mm*) or O (*Amm*2) space group led to poor fitting results with high R-factors. A thorough observation of the x-ray diagrams of these compositions that present enlarged peaks and taking into account the dielectric broadened anomaly around room temperature (not classically λ-shape), we concluded in favor of coexistence of symmetry phases. As reported previously in Sn doped BT materials [15], the phase diagram in this family shows the existence of a quasi-quadruple point that depends on the doping element. Indeed, a two-phase structural model (*P4mm* and *Amm*2) has been tested together in accordance to phase diagram prediction that leads to a satisfactory profile fitting and Rietveld calculation compared to that of single-phase structural models.

Similarly, the structural analysis of x-ray data for the composition x = 0.10 requires a structural refinement based on a mixture of T, O and R (*R3m*) phases together to attempt realistic results. The calculated lattice parameters, reliability factors and phase fractions are gathered in Table 1. From this Table 1, a relatively high value of $R_{wp} \sim 11$ is observed for our samples. According to Toby [16], the relatively high values of the R-factors can be attributed to the conditions of the data collection such as high counting statistics and background levels. Similar results have been found in similar systems $BaTi_{1-x}Sn_xO_3$ [17, 18]. Indeed, the most important way to determine the quality of a Rietveld fit is to check the observed and calculated patterns graphically and to ensure that the model is chemically plausible [16]. The quality of the Rietveld refinement is depicted in Fig. 2 that shows the calculated diagram, observed one and their difference accordantly. These structural analysis results corroborate microstructural observations, larger grain size and compactness ceramics seems to be evidenced in compound with x = 0.05, 0.07 and 0.10.

**(Insert Fig. 2, here)**

**(Insert Table 1, here)**

### 3.2 Raman Spectroscopy

Raman spectroscopy is a powerful tool that provides a very sensitive measurement of the local crystal symmetry in single crystals, polycrystals and powders [19–21]. Unpolarized Raman spectra of BSTS-x ceramics recorded at RT are shown in Fig. 3. $BaTiO_3$ possesses at RT a perovskite structure with 5 atoms per primitive cell. Thus, in its C phase, there are 12 optical vibrational active modes, belonging to $3F_{1u}+F_{2u}$ irreducible representations. The $F_{1u}$ modes are infrared active while the $F_{2u}$ mode is called "silent" mode since it is neither infrared nor Raman active. When the phase structure transforms into the T phase, the $F_{1u}$ mode splits into $A_1$ and E modes, while the $F_{2u}$ splits into $B_1$ and E modes. The $A_1$ and E modes are both Raman and infrared active, while $B_1$ is only Raman active. Moreover, the $A_1$ and E mode further split into transverse (TO) and longitudinal (LO) optical modes when an electrostatic force is induced. When the phase structure changes from T to O, the optical modes belong to the symmetrical $A_1$, $A_2$, $B_1$, and $B_2$ modes that are Raman active. In the R phase, there are $A_1$ and E modes, which are also Raman active [22].

As shown in Fig. 3, all Raman active modes of $A_1$(TO) are observed as a broad bands at 238 cm$^{-1}$ and 516 cm$^{-1}$, $B_1$/E(TO, LO) phonon mode is observed as a sharp peak at 302 cm$^{-1}$ and $A_1$/E(LO) mode of low intensity can be found at 724 cm$^{-1}$. Those Raman active modes are typical in the Raman spectrum tetragonal phase of BT, as reported in the literature [17, 20]. For x = 0.00 and x = 0.02, the presence of an interference effect at 179 cm$^{-1}$, which occurs only in the T phase [20], the sharp peak at 302 cm$^{-1}$ and asymmetric bands at 238, 516 and 724 cm$^{-1}$ are the signature of T symmetry [12, 20]. As x increases, the peak marked by an arrow (Fig. 3) appears near 186 cm$^{-1}$ and the band $A_1$(TO) ~ 238 cm$^{-1}$ becomes flat, implying that the phase transition between O phase and T phase occurs for $0.05 \leq x \leq 0.07$. Besides, the strong peak characteristic of the Rhombohedral phase appear near 186 cm$^{-1}$, as well as the peak at 238 cm$^{-1}$ for x = 0.10 [23]. Due to its similarity to O phase at this composition, R, O and T phases coexist at x = 0.10. However, the reduced sharpness of the peak at 302 cm$^{-1}$ and the dip that becomes indistinct at 179 cm$^{-1}$ for x = 0.10 suggest that the tetragonal phase is not dominant at this composition. On further increasing x to 0.15, the bands situated at 302 cm$^{-1}$ and 724 cm$^{-1}$ disappear. In comparison, the one located at around 238 cm$^{-1}$ becomes broad and flat, revealing that the phase structure is changed into the paraelectric C phase, as already reported in previous works [12, 17, 20]. These results are in accordance with XRD (section 3.1) and the dielectric results presented in the next section.

**(Insert Fig. 3, here)**

### 3.3 Microstructure analysis

The microstructural features of BSTS-x ceramics sintered at their optimized temperatures are shown in Fig. 4 (a-f). All samples showed good compactness due to the chosen high sintering temperatures. For samples with x increasing from 0.00 up to 0.07, the micrographs show homogeneous grain distribution and clear grain boundaries that confirm the enhancement of the ceramics' density for excellent electrical properties [24]. Besides, at x = 0.10, the ceramic displays an inhomogeneous grain distribution with big grains surrounded by many smaller ones, which is probably related to the coexistence of phases at a tricritical point and discontinuous grain growth [25]. In the sample with x = 0.15, the grain growth is inhibited leading to the average grain size decrease from ~68 µm at x = 0.02 to ~35 µm at x = 0.15, in good agreement with previous results [25, 26]. The reason for such large grain sizes for doped BSTS-x ceramics could be linked to higher sintering temperature adopted for densification. The density of ceramics was measured based on Archimedes' principle and acquire a relative density samples in the range of 95-99 % of theoretical density.

**(Insert Fig. 4, here)**

### 3.4 Dielectric analysis and the phase diagram

To further clarify the phase transition in the BSTS-x system, temperature dependence of the dielectric constant $\varepsilon'$ for all studied BSTS-x ceramics at 1 kHz is gathered in Fig. 5. Three dielectric peaks corresponding to the phase transitions (R-O, O-T and T-C or R-C) denoted as $T_{R-O}$, $T_{O-T}$ and $T_C$, respectively, are observed for x < 0.10. With increasing x, the $T_C$ decreases, while $T_{R-O}$ and $T_{O-T}$ are shifted up toward RT. Consequently, $T_{R-O}$, $T_{O-T}$ and $T_C$ are evidenced by $\partial \varepsilon'/\partial T$ near RT for x = 0.07, as depicted in the inset of Fig. 5. With increasing x further to 0.15, all three transitions merge, forming one broad peak at $T_C$. A more in-depth analysis of the dielectric results permit to confirm the merging of the R–O, O–T and R-C phase transitions at x = 0.10, i.e., formation of the quasi-quadruple point near the RT in the phase diagram. The above result is well consistent with our XRD and Raman study. Also, only one dielectric peak characteristic of ferroelectric to paraelectric phase transition (T-C or R-C) is observed for x ≥ 0.10. A similar merging tendency of dielectric anomalies is also found in the temperature dependence of the loss factor $tg\delta$ (Fig. 6). The values of $T_{R-O}$, $T_{O-T}$, $T_C$, and some other dielectric parameters were summarized in Table 2.

As shown in Fig. 5, a significant increase of the maximal dielectric permittivity ($\varepsilon'_m$) is observed for Sn-compositions up to x = 0.10, this increase is followed by a drastic drop for compositions approaching x = 0.15. Note that ($\varepsilon'_m$) corresponds to the F-P transition at $T_C$, which itself decreases when Sn-amount increases. This behavior is similar to previously reported results on Sn-substituted BaTiO$_3$ [27]. Shi *et al*. described the enhancement of the dielectric properties for Sn doped BT [28]. They have found that, as the concentration of Sn increases in the material, the chemical distribution of Sn is more inhomogeneous. Consequently, already formed Polar Nano-Regions (PNRs), which should have smaller lattice constants due to their lower Sn concentration, expand under the stress of the matrix [28]. Therefore, the dielectric constant values are greatly enhanced [28].

**(Insert Fig. 5, here)**

**(Insert Fig. 6, here)**

**(Insert Table 2, here)**

The phase diagram of BSTS-x ceramics (Fig. 7) was established using the dielectric measurements. Three ferroelectric regions related to R, O and T symmetry phases and one paraelectric C symmetry phase are surrounding the quasi-quadruple point (QP) at x = 0.10 in the diagram. It can be seen that $T_C$ decreases linearly, whereas both $T_{R-O}$, $T_{O-T}$ shift to approach RT. As a result, a QP is formed at the composition x = 0.10 at RT in accordance with XRD and Raman spectrum analysis. As mentioned above, the quasi-quadruple point describes a state of four-phase coexistence (R-O-T-C) that may exist in our system, according to generalized Gibbs rule [29]. The possible existence of such multiphase points have been found in similar systems BaTi$_{1-x}$Sn$_x$O$_3$ [30, 31], SrTiO$_3$ [32], PbZr$_{1-x}$Ti$_x$O$_3$ [33], Eu$_x$Sr$_{1-x}$TiO$_3$ [34], and Ba(Zr$_x$Ti$_{1-x}$)O$_3$ [35].

**(Insert Fig. 7, here)**

### 3.5 Ferroelectric analysis

The temperature-dependent *P–E* hysteresis loops recorded at the frequency of 5 Hz are shown in Fig. 8(a). Typical ferroelectric *P-E* hysteresis loops are observed for all the samples. Notably, a slimmer hysteresis loop is obtained with increasing temperature, which transforms into a linear response that attests ferroelectric-to-paraelectric phase transition in every composition. The evolution of the spontaneous polarization ($P_s$) and the coercive field ($E_C$) as a function of Sn-content at RT is displayed in Fig. 8(b). As x increases, the $P_s$ slightly decreases first at local minimum for x = 0.02 and then increases and reaches a maximum value of 12.93 µC/cm² at x = 0.05. For compositions above x = 0.05 it again decreases gradually to very low value below 3 µC/cm² for x = 0.15. This behavior shows different dipole orientations through the corresponding ferroelectric phases when x increases. However, the coercive field $E_C$ seems to have a particularly low value desirable for energy storage applications. $E_C$ decreases from 1.20 kV/cm at x = 0.00 to 0.22 kV/cm at x = 0.10 and even approaches zero at x = 0.15. These results confirm that the composition with x = 0.15 is already in its paraelectric cubic phase at RT as confirmed by XRD, Raman spectroscopy and dielectric measurements (with $T_C$ = -13 °C).

**(Insert Fig. 8, here)**

DC electric-field dependence of the dielectric constant of BSTS-x ceramics recorded at RT and frequency of 1 kHz is plotted in Fig. 9 (a). The typical butterfly shape of the curves for samples with x ranges from 0.00 to 0.10 confirms the ferroelectric nature of these ceramics. No butterfly behavior was obtained for the composition x = 0.15 as expected for its cubic paraelectric symmetry phase. It is demonstrated that the dielectric constant gets rapidly suppressed when the electric field is applied. In particular, at x = 0.10, only 1.40 kV/cm is required to suppress $\varepsilon'$ from its highest value of ~ 35000 to ~ 12000, which is suppression of more than 60% of its original value. These results show high dielectric tunability in bulk BSTS-x. The relative tunability (η) is usually calculated from $\varepsilon - E$ curve according to the following expression [6] :

$$\eta = \frac{\varepsilon'(0) - \varepsilon'(E)}{\varepsilon'(0)} \times 100. \tag{1}$$

Here $\varepsilon'(0)$ and $\varepsilon'(E)$ represent the dielectric constant value at an applied electric field zero and E, respectively. The evolution of the relative tunability as a function of the composition x is displayed in Fig. 9 (b). Tunability values lower than 15% are observed for $0.00 \leq x \leq 0.05$, at x = 0.07 the η increases to around 30% and finally reaches its maximum value of 63% at x = 0.10 under a low bias electric field of 1.40 kV/cm. This behavior is in accordance with *P-E* hysteresis loops, which show low Ec and very slim *P-E* hysteresis loops at x = 0.10 near its F-P phase transition at RT (Fig. 8(a)). Therefore, BSTS-10 with a giant dielectric tunability at RT is very competitive to the currently studied tunable ferroelectrics [11, 36–38]. Moreover, the applied bias electric field of 1.40 kV/cm in this study is very low, and relative tunability could increases further under higher electric field values.

It is worth recalling that for tunable device applications, both high tunability and low dielectric losses are needed. The dielectric loss in the studied samples, especially for x = 0.10, seems to be average (around 0.06 at 0kV/cm) near RT (inset of Fig. 6). This result appears to render BSTS-x almost unfavorable for tunable applications at RT. However, the relatively high value of $tg\delta$ in bulk BSTS-x ceramics can be further reduced by additives to this solid solution and $\varepsilon'$ can be tailored to the appropriate values for tunable practical applications.

**(Insert Fig. 9, here)**

## 4. Conclusions

In the present paper, $Ba_{0.9}Sr_{0.1}Ti_{1-x}Sn_xO_3$ sintered ceramics were studied for different compositions (x = 0.00 to 0.15). The results show that increase of Sn-amount induces the following phase transformation: the crystal structure for x = 0.00 and x = 0.02 shows best agreement with T (*P4mm*) phase; for x = 0.05 and x = 0.07 the analysis showed coexistence of T (*P4mm*) – O (*Amm*2) phases, while for x = 0.10 coexistence of T (*P4mm*) – O (*Amm*2) – R (*R3m*) phases. The composition with x = 0.15 showed undistorted C (*Pm-3m*) phase. By Raman spectroscopy, we confirmed the XRD results and dielectric analysis. It is found that with increasing Sn-amount in BST lattice, the three transitions merge into one broad peak near RT at x = 0.10, i.e., at the quasi-quadruple point in the phase diagram. As a new result, the dielectric constant is greatly improved ($\varepsilon' \sim 35000$ at x = 0.10). Temperature-dependent *P–E* hysteresis loops confirmed ferroelectric behavior for the compositions from x = 0.00 to x = 0.10 and paraelectric linear behavior for the composition x = 0.15 in accordance to dielectric and XRD results. DC-electric-field dependence of the dielectric constant at RT revealed a relative tunability of 63% obtained at x = 0.10 under low bias electric field of 1.40 kV/cm. Such results make BSTS-x ceramics with x = 0.10 very promising potential matrix for application in electrically tunable devices at RT.

## Acknowledgements

The authors are grateful to the financial support received from the European Community through the H2020MSCA-RISE-ENGIMA-778072 project. The support from ARRS project J1-9147, program P1-0125 and the CNRST Priority Program (PPR15/2015) is gratefully acknowledged.**References**

[1] A. Ioachim *et al.*, "Barium strontium titanate-based perovskite materials for microwave applications," *Prog. Solid State Chem.*, vol. 35, pp. 513–520, 2007, doi: 10.1016/j.progsolidstchem.2007.01.017.

[2] J. Wang, X. Yao, and L. Zhang, "Preparation and dielectric properties of barium strontium titanate glass-ceramics sintered from sol – gel-derived powders," *Ceram. Int.*, vol. 30, pp. 1749–1752, 2004, doi: 10.1016/j.ceramint.2003.12.134.

[3] C. Mao, X. Dong, T. Zeng, H. Chen, and F. Cao, "Nonhydrolytic sol – gel synthesis and dielectric properties of ultrafine-grained and homogenized $Ba_{0.70}Sr_{0.30}TiO_3$," *Ceram. Int.*, vol. 34, pp. 45–49, 2008, doi: 10.1016/j.ceramint.2006.08.002.

[4] J. Cheng, J. Tang, J. Chu, and A. Zhang, "Pyroelectric properties in sol – gel derived barium strontium titanate thin films using a highly diluted precursor solution," *Appl. Phys. Lett.*, vol. 77, no. 7, pp. 1035–1037, 2000, doi: 10.1063/1.1289038.

[5] A. Kumar and S. G. Manavalan, "Characterization of barium strontium titanate thin films for tunable microwave and DRAM applications," *Surf. Coat. Technol.*, vol. 198, pp. 406–413, 2005, doi: 10.1016/j.surfcoat.2004.10.044.

[6] A. K. Tagantsev, V. O. Sherman, K. F. Astafiev, J. Venkatesh, and N. Setter, "Ferroelectric materials for microwave tunable applications," *J. Electroceramics*, vol. 11, no. 1–2, pp. 5–66, 2003, doi: 10.1142/s2010135x12300022.

[7] Q. Zhang, J. Zhai, Q. Ben, X. Yu, and X. Yao, "Enhanced microwave dielectric properties of $Ba_{0.4}Sr_{0.6}TiO_3$ ceramics doping by metal Fe powders," *J. Appl. Phys.*, vol. 112, no. 10, pp. 0–10, 2012, doi: 10.1063/1.4766276.

[8] N. Baskaran and H. U. A. Chang, "Effect of Sn doping on the phase transformation properties of ferroelectric $BaTiO_3$," *J. Mater. Sci. Mater. Electron.*, vol. 2, pp. 527–531,12

**Tables and figures captions**

Table 1: Lattice parameters, cell volume and reliability factors of BSTS-x ceramics.

Table 2: Dielectric parameters of BSTS-x ceramics.

Figure 1: Room temperature XRD patterns of BSTS-x ceramics in the 2θ of (a) 10° – 80°, and (b) 44° – 46°.

Figure 2: Rietveld refined XRD patterns of BSTS-x ceramics recorded at RT.

Figure 3: Room temperature Raman spectra of BSTS-x ceramics (x = 0 – 0.15).

Figure 4: SEM micrographs of BSTS-x ceramics for various Sn-amounts.

Figure 5: Temperature dependence of $\varepsilon'$ for all the compositions x at 1 kHz of frequency. Inset displays derivative of $\varepsilon'$ versus temperature ($\partial \varepsilon'/\partial T$) for the composition x = 0.07.

Figure 6: Temperature dependence of $tg\delta$ for all the compositions x at 1 kHz of frequency. The inset shows the temperature and frequency dependence of $tg\delta$ for x = 0.10.

Figure 7: Phase diagram of BSTS-x ceramics (x = 0 – 0.15).

Figure 8: (a) Temperature dependence *P-E* hysteresis loops at 5 Hz of BSTS-x specimens (x = 0.10). (b) Variation of $P_s$ and $E_C$ as a function of the composition x (x = 0 – 0.15).

Figure 9: (a) Dielectric constant versus electric field (*ε–E*) characteristics of BSTS-x ceramics (x = 0 – 0.10) recorded at RT. (b) Evolution of the relative tunability versus the composition (x) of BSTS-x ceramics.

Table 1

| x | Space group | Unit Cell parameters | | | | Phase fraction (%) | R-factors | | | | |
|---|---|---|---|---|---|---|---|---|---|---|---|
| | | a(Å) | b(Å) | c(Å) | V(Å³) | | $\chi^2$ | $R_p$ | $R_{wp}$ | $R_B$ | $R_F$ |
| **0.00** | P4mm | 3.9857 | - | 4.0160 | 63.797 | 100 | 1.979 | 10.2 | 14.1 | 2.69 | 3.85 |
| **0.02** | P4mm | 3.9912 | - | 4.0144 | 63.948 | 100 | 1.516 | 10.5 | 13.5 | 5.40 | 7.23 |
| **0.05** | P4mm | 3.9966 | - | 4.0122 | 64.085 | 54.25 | 1.164 | 8.79 | 11.7 | 4.02 | 5.58 |
| | + | | | | | | | | | | |
| | Amm2 | 3.9968 | 4.0071 | 4.0006 | 64.073 | 45.75 | | | | 4.27 | 5.74 |
| **0.07** | P4mm | 4.0001 | - | 4.0077 | 64.126 | 64.89 | 1.361 | 9.20 | 12.5 | 3.53 | 4.56 |
| | + | | | | | | | | | | |
| | Amm2 | 3.9937 | 4.0004 | 4.0143 | 64.134 | 35.11 | | | | 3.07 | 3.87 |
| **0.10** | P4mm | 4.0011 | - | 4.0103 | 64.200 | 49.09 | 1.981 | 6.65 | 8.30 | 1.87 | 2.41 |
| | + | | | | | | | | | | |
| | Amm2 | 3.9944 | 4.0204 | 4.0128 | 64.441 | 12.24 | | | | 4.79 | 6.41 |
| | + | | | | | | | | | | |
| | R3m | 4.0125 | - | - | 64.604 | 38.66 | | | | 1.52 | 2.14 |
| **0.15** | Pm-3m | 4.0073 | - | - | 64.352 | 100 | 1.061 | 8.61 | 11.4 | 2.28 | 3.51 |

Table 2

| Composition x | 0.00 | 0.02 | 0.05 | 0.07 | 0.10 | 0.15 |
|---|---|---|---|---|---|---|
| $T_{R-O}$ (°C) | -69 | -65 | -23 | -2 | - | - |
| $T_{O-T}$ (°C) | 2 | 5 | 14 | 23 | - | - |
| $T_C$ (°C) | 95 | 86 | 62 | 50 | 26 | -13 |
| $\varepsilon'_{max}$ | 14582 | 17087 | 18828 | 29078 | 33187 | 19360 |
| $\varepsilon'$ (RT) | 3313 | 3162 | 6346 | 12186 | 32744 | 5942 |
| $tg\delta$ ($T_C$) | 0.017 | 0.064 | 0.051 | 0.049 | 0.062 | 0.028 |

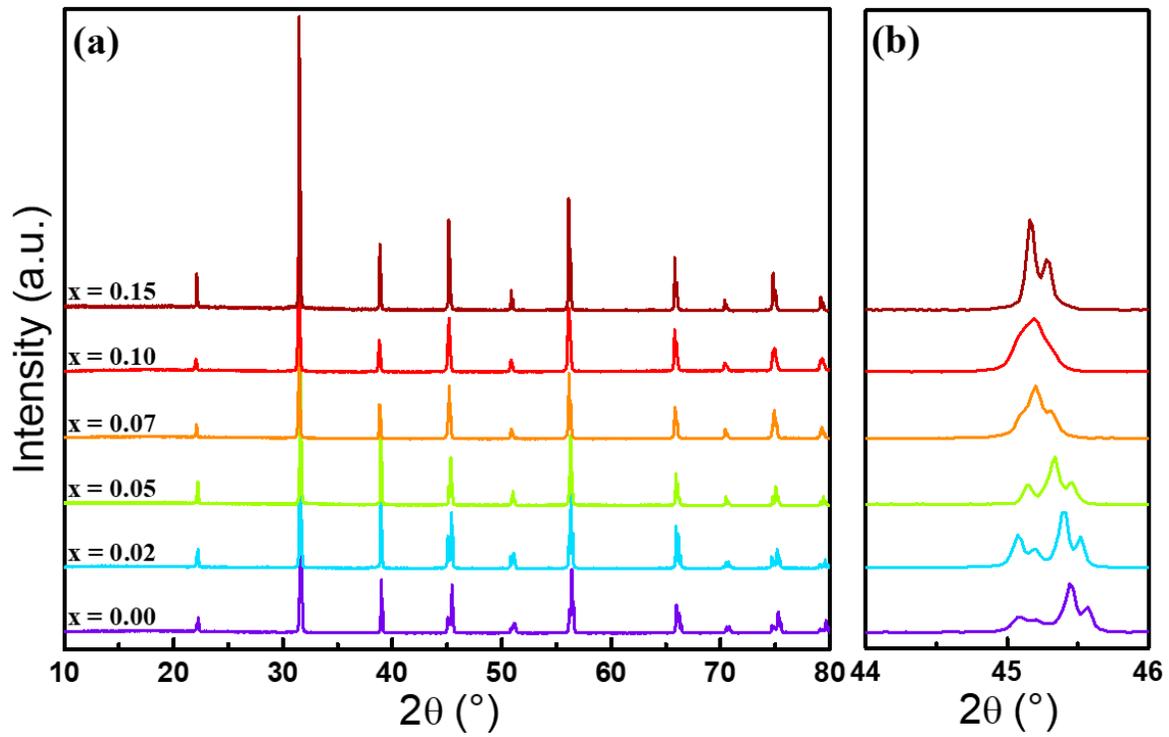

Figure 1

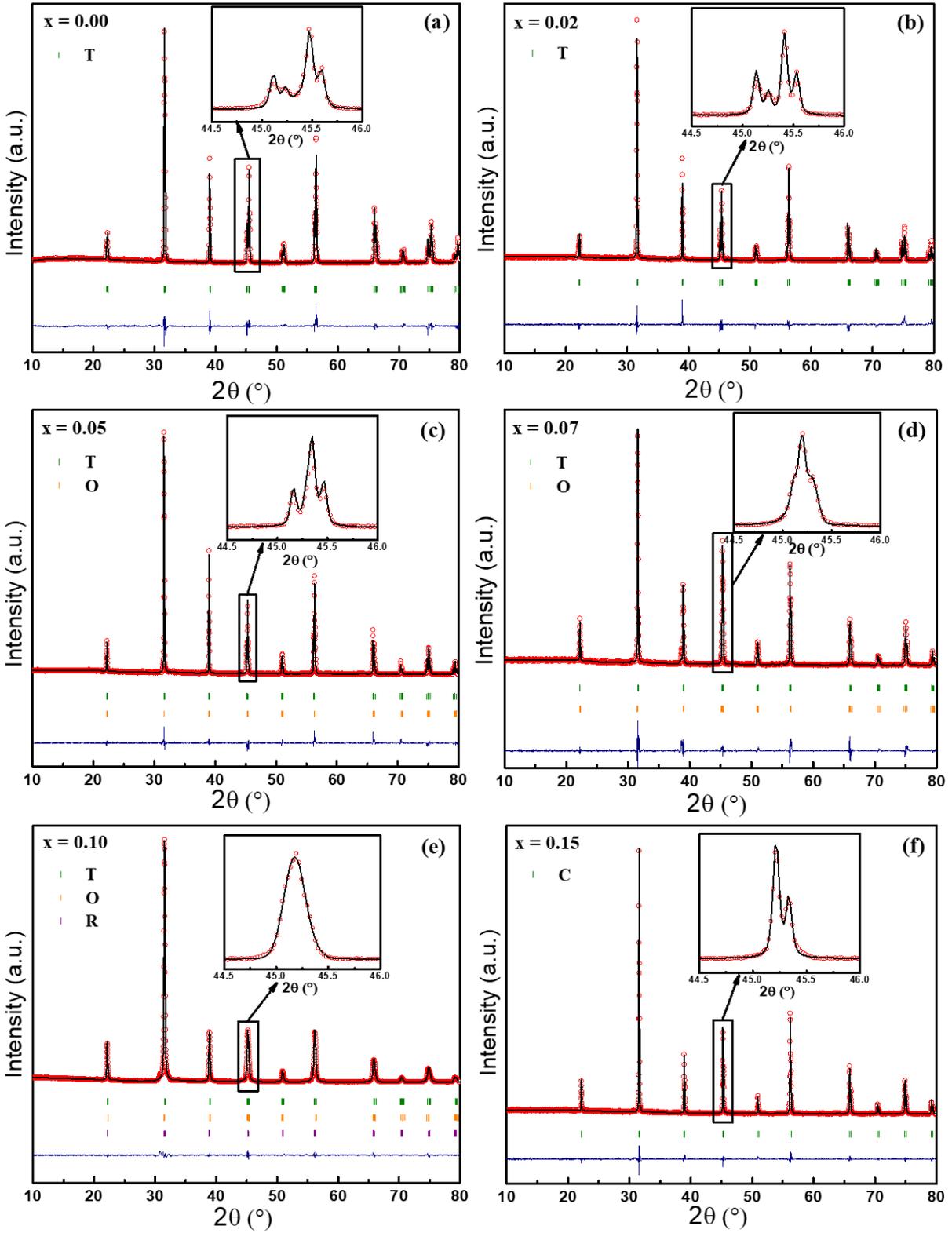

Figure 2

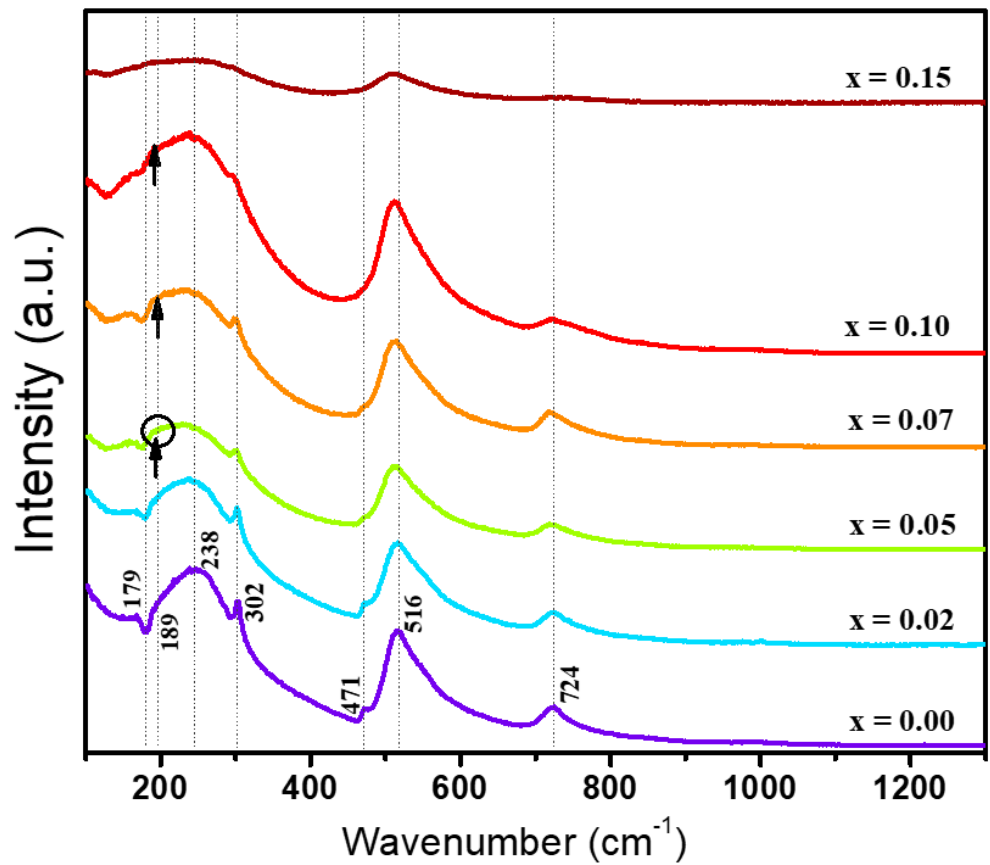

Figure 3

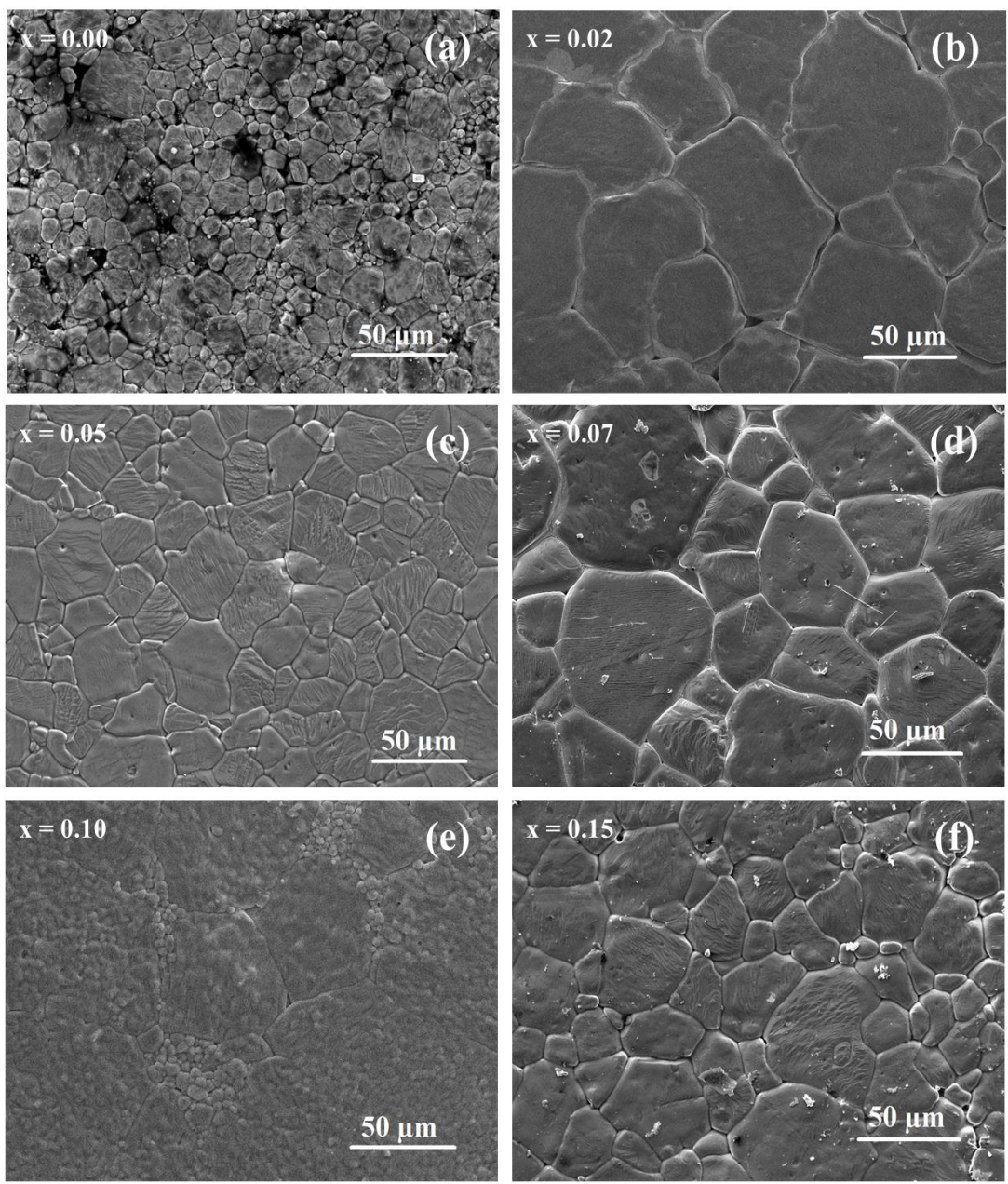

Figure 4

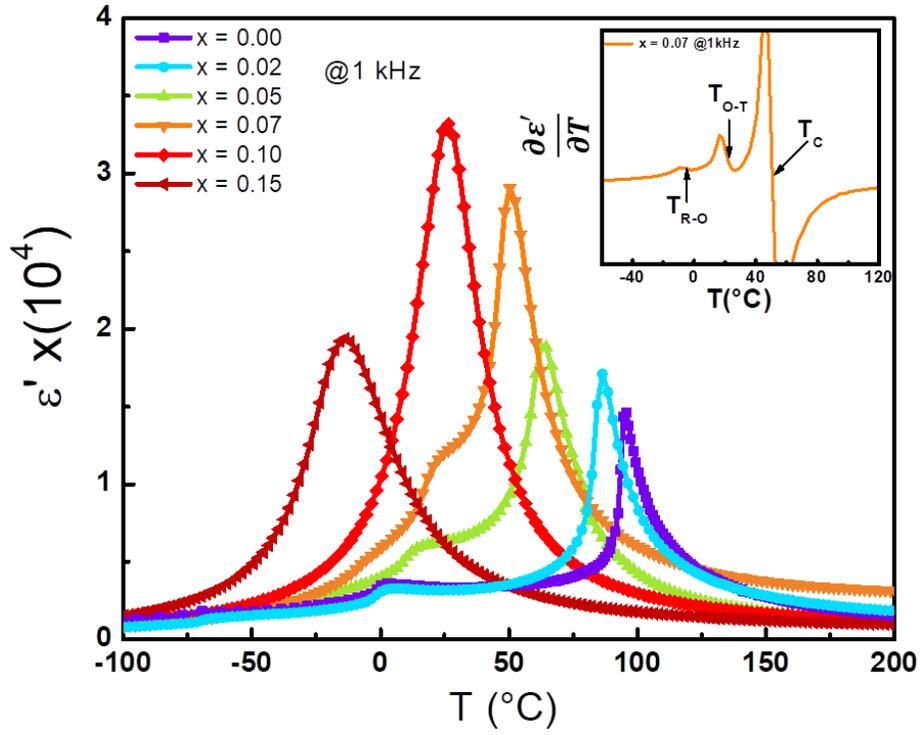

Figure 5

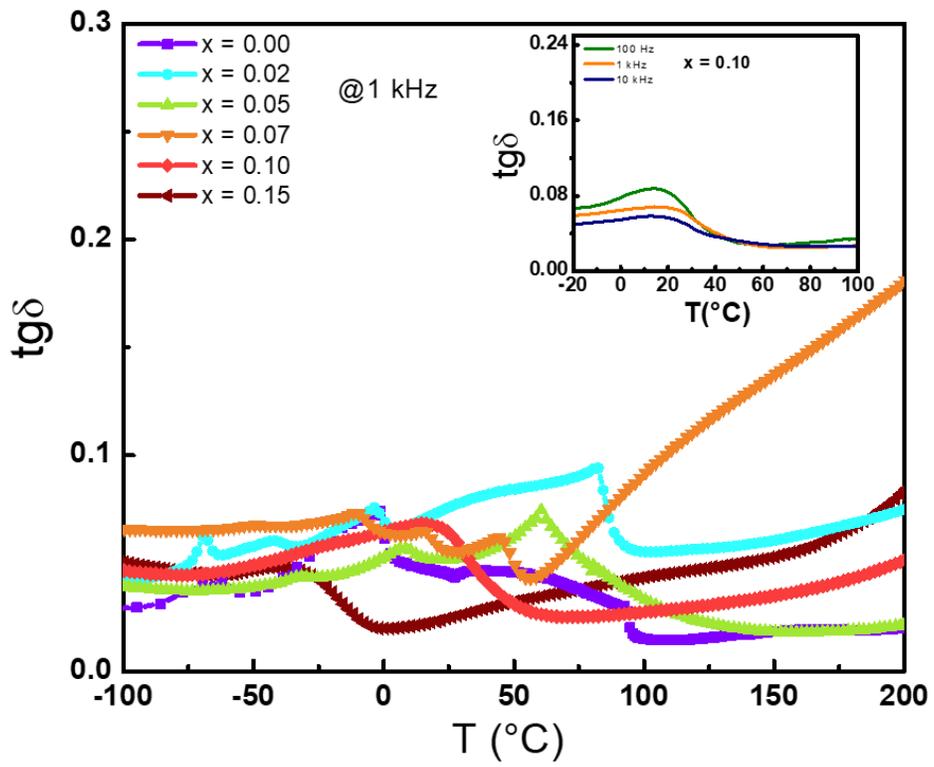

Figure 6

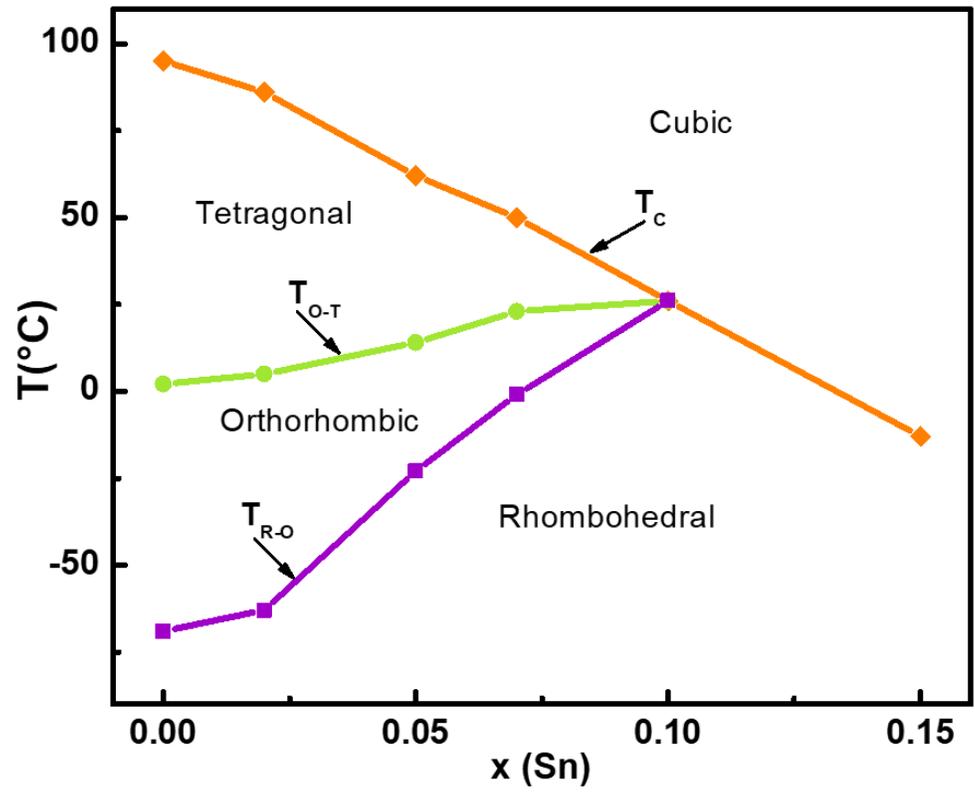

Figure 7

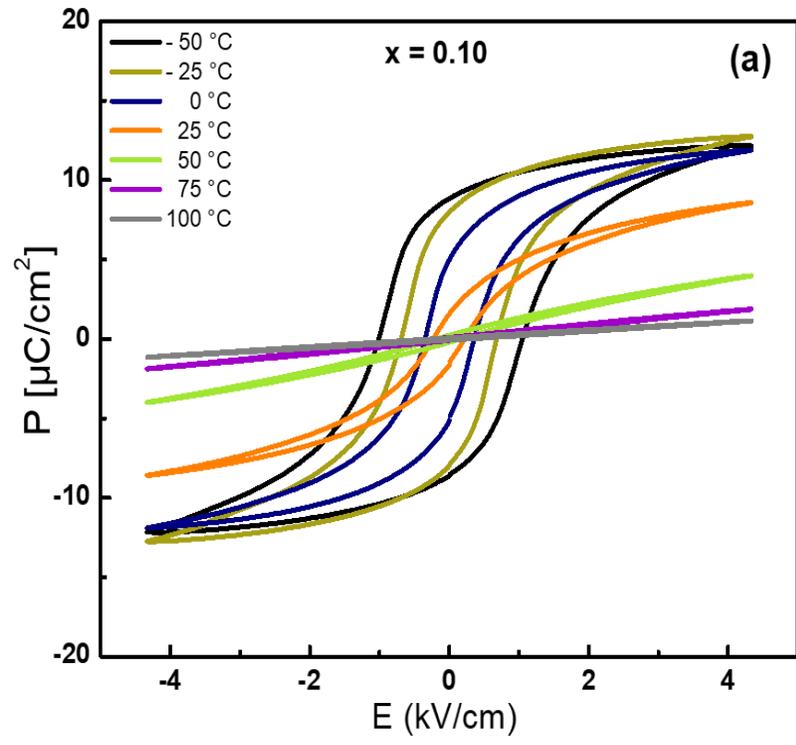
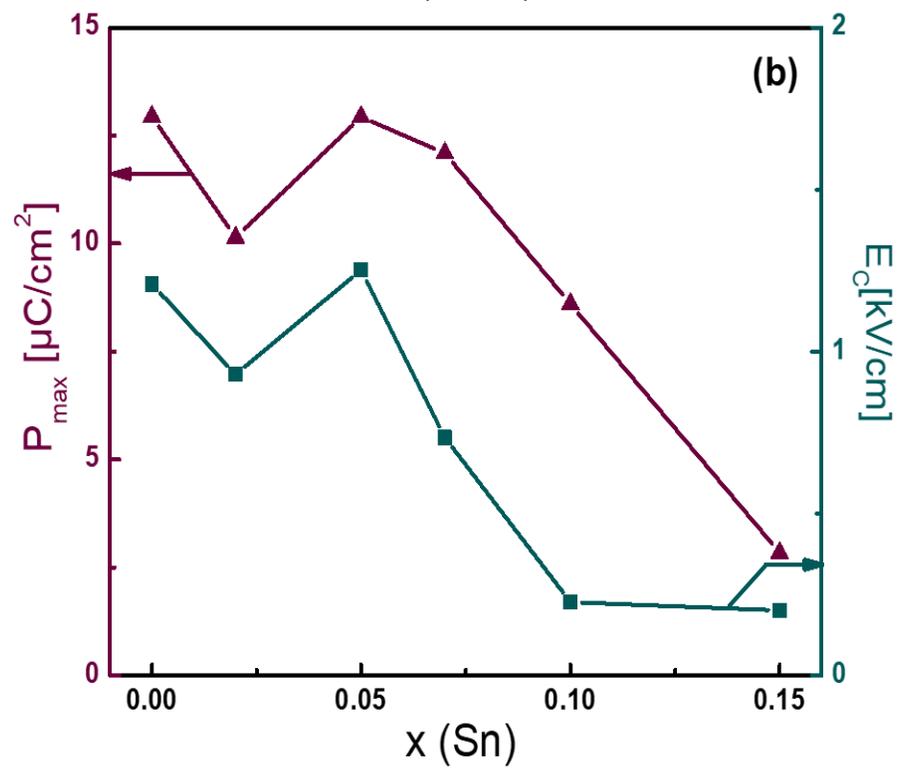

Figure 8

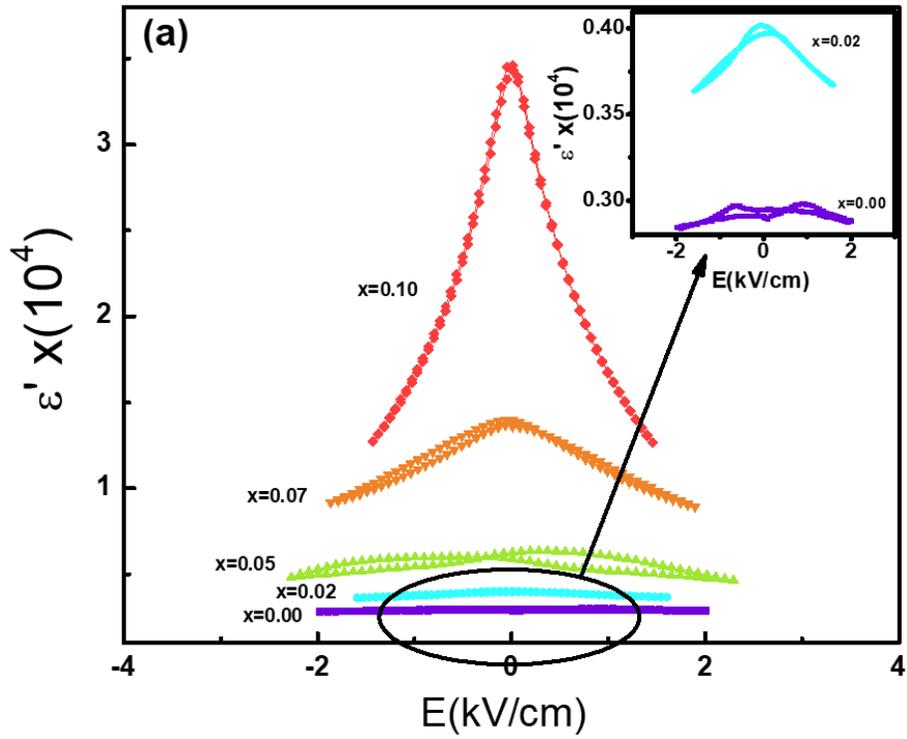
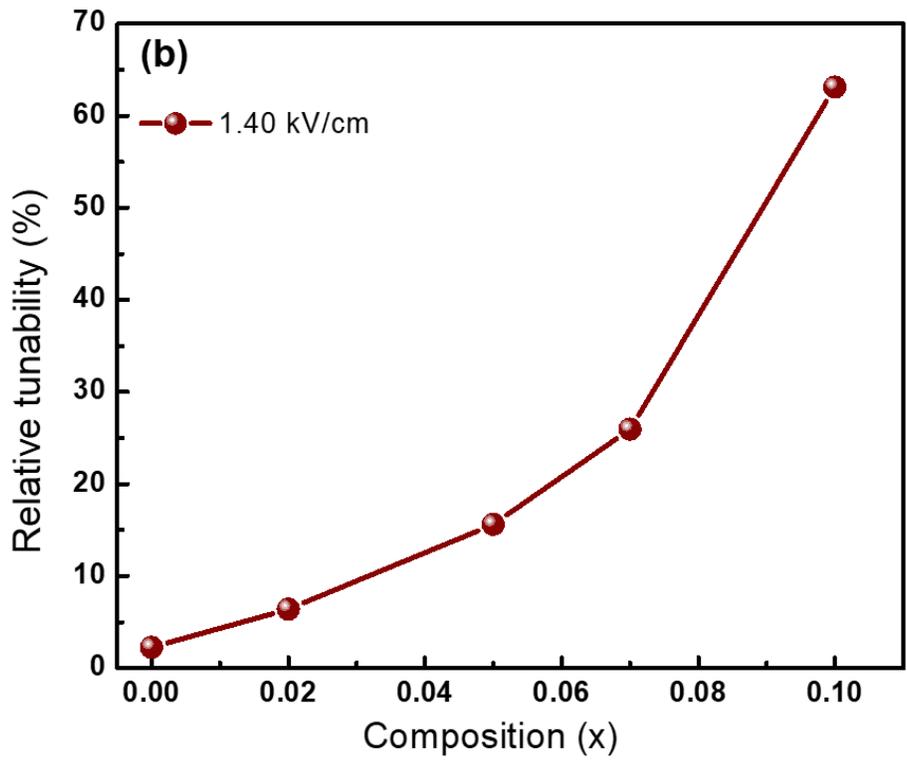

Figure 9